\documentclass[twocolumn]{article}
\usepackage{amsmath,amssymb}

\input xy
\xyoption{all}
\xyoption{arc}
\usepackage[curve]{xy}
\xyoption{poly}

\begin{document}
\title{Dirac's Large Number Hypothesis\\
and Quantized Friedman Cosmologies}
\author{ James G. Gilson\quad  j.g.gilson@qmul.ac.uk \thanks {
 School of Mathematical Sciences,
Queen Mary University of London, Mile End Road, London E1 4NS,
United Kingdom.}}

\maketitle

\begin{abstract}

This paper contains applications of the quantum theory for gravity developed in the paper {\it A Sketch for a Quantum Theory of Gravity\/}.  Firstly, it is shown that the theory gives a direct derivation of the implications of Dirac's large number hypothesis. The three dimensionless large numbers are obtained as  three closed formulae with definite coefficients constructed from known physical constants. Secondly, the theory is used to derive two quantum Friedman cosmologies. The first of these Cosmologies is a very simple first approximation involving a key angular parameter being zero and the second is a more accurate one involving the same key parameter not being zero. The cosmological constant plays a basic and fundamental part in the formalism with very accurate agreement with measurement pertaining.
\end{abstract}
\section{The Three Large Numbers}
\setcounter{equation}{0}
\label{sec-ttln}
The paper {\it A Sketch for a Quantum Theory of Gravity}\cite{Gil:2005}, will referred to as $A$ in this paper.
The three large numbers that appeared in the works of Dirac\cite{Dir:1974}, Eddington\cite{Edd:1946}  and others\cite{Kil:1994}  have been extensively discussed their meanings, possible relations between them and their significance has been studied in great detail by many workers\cite{Mis:1973}. Many attempts have been made to explain their physical philosophical position in the natural world. Generally they have been regarded as presenting us with a great mystery that needs to be resolved, if we are ever to understand the relation between the very small world of atoms and the very large cosmological world. The quantum-gravitation theory developed in $A$ does explain just how and why they occur and how they are related. The noted three large numbers are the numerical values of the quantities that are often called  {\it the relative strength of the electrical and gravitational forces between an electron and a proton, $\xi$ \/}; {\it the ratio of the length scale associated with the universe, $c/H$\/}, \cite{Nar:1993}, and {\it the classical radius of the electron, $r_e$\/}; and lastly, {\it N, the number of particles in the universe\/}. They are listed below in the order just mentioned. 
\begin{eqnarray}
\xi& = & \alpha\hbar c/G m_pm_e \approx 2.27\times 10^{39}\nonumber\\
& \approx & 10^{40}\ !\label{1.01}\\
(c/H)/r_e& \approx & 10^{40}\label{1.02}\\
N& \approx & 10^{80}\label{1.03}
\end{eqnarray}
where $H$ is Hubble's constant. The first of these $\xi$ can be calculated from the given formula in the first entry to equation (\ref{1.01}) and on the basis of present day measured values for the constants $\alpha$, $m_p$, $m_e$ and $G$ it assumes a value of order $10^{39}$. The first three of these constants are from quantum measurements and are known very accurately. The fourth, $G$ is from astronomical and other macroscopic measurements and its value is not known to anything like the same order of accuracy as the quantum constants. Nevertheless the calculated value of $\xi$ is the value cited on the first line of equation (\ref{1.01}). I can only assume that the suggested order of magnitude often quoted, $10^{40}$, and displayed in the second line with a $!$ is wishful thinking in attempting to see a close relation between $\xi$ and $(c/H)/r_e$.
There seems little reason to expect these quantities to be equal just because they have the same large order of magnitude. On the same tack, I cannot see that there is any reason to expect that $N$ is the numerical square of $(c/H)/r_e$. However, it will next be shown that the orders of magnitude of these quantities are derivable from the quantum-gravitation theory in $A$. Thus demonstrating that their relationships are indeed physical and not accidental as some workers believe. Dirac's large number hypothesis, LNH, is essentially the conclusion that these three large number valued relations between quantum and cosmological quantities are not accidental. They rather represent relations that hold for all time throughout the whole history of the universe. On the main interpretation of (LNH), it is implied  that $G$ must vary inversely with epoch. Using the quantity $\xi$, we get the approximate value of $N_G$ from experimental numerical values together with gravitation theory as
 
\begin{eqnarray}
N_G & = & \alpha ^{-1}\xi \cos (\chi _G(N_G)) \label{1.04}\\
  & \approx & 3.11 \times 10^{41}. \label{1.05}\\
 \xi & = & \alpha N_G/\cos (\chi _G(N_G)). \label{1.06}
\end{eqnarray}

The factor $\cos (\chi _G(N_G))$ on the right hand side of equations (\ref{1.04}) and (\ref{1.06}) does not detract from the evaluation of $N_G$ because it is so close to unity as to be measurably indistinguishable from it. This is because of the largeness of $N_G$. Equation (\ref{1.04}) is just the inverse of equation (\ref{1.06}) under the same understanding about the factor $\cos (\chi _G(N_G))$. It shows how the observed large value order of $\xi$ is determined by the key state integer or eigen-number, $N_G$.
Let us now consider the second of the three observed large numbers $(c/H)/r_e$. Within the theory we have, from equations (3.22$A$) and (2.5$A$), its definition is,
\begin{eqnarray}
(c/H^*)/r_e & = & m_ec^2t^*/(\alpha\hbar) \label{1.07}\\
  & = & \frac{m_eN_G}{\alpha m_p \cos ^3 (\chi _G(N_G))} \nonumber\\
& = & 3.36\times 10^{40}\label{1.08}\\
\frac{m_e}{\alpha m_p \cos ^3 (\chi _G(N_G))}&\approx & 7.46\times 10^{-2}\approx 10^{-1}\nonumber\\
& & \label{1.09} 
\end{eqnarray}

Thus here again we find that the largeness of $(c/H)/r_e$ is determined by the key quantum integer $N_G$, the order of largeness of $(c/H)/r_e$ is brought down into line with $\xi$ by the relatively small factor in equation (\ref{1.09}).
Let us now consider the last large number $N$, the number of particles in the universe. This is related to the mass of the universe $M_U$ from the theory and is given by equation (3.4$A$),
\begin{eqnarray}
M_U & = & N_G m_e\alpha ^{-1}_G\label{1.10}\\
& = & \frac{N_G^2 m_e}{\cos (\chi_G(N_G))} \label{1.11}\\
M_U/m_p & = & \frac{N_G^2 m_e}{ m_p \cos (\chi_G(N_G))}\nonumber\\
&\approx & 5.27\times 10^{79}\nonumber\\
&\approx & 10^{80}\label{1.12}\\
\frac{m_e}{ m_p \cos (\chi_G(N_G))}& \approx & 5.4\times 10^{-4}.\label{1.13}
\end{eqnarray}
Collecting the three results together with the best value that can be assessed from theory and experiment for the current value of $N_G$, we have
\begin{eqnarray}
\xi & = & \frac{\alpha N_G}{\cos (\chi_G(N_G)}\label{1.14}\\
(c/H)/r_e& = & \frac{ m_e N_G }{\alpha m_p \cos ^3 (\chi _G(N_G))} \label{1.15}\\
M_U/m_p & = & \frac{ m_e N_G^2 }{ m_p \cos (\chi_G(N_G))}=N_p\label{1.16}\\
N_G & \approx & 3.11 \times 10^{41}. \label{1.17}
\end{eqnarray}
Here also we get the clear representation for the number $N_p$ of {\it protonic\/} sized rest masses in the universe as determined by the square of the key state quantum number $N_G$.

All three large numbers can thus be identified as definite functions of the microscopic  constants $\alpha$, $m_e$ and $m_p$ together with $c$ and the key quantum cosmological state quantum number $N_G$. This state parameter is the source of the large numbers and the basis of Dirac's large number hypothesis.

The model quantum system developed in $A$ conforms exactly to the Dirac large number hypothesis and at the same time fills in the detailed construction for the quantum theory necessary to fulfil that role. All the coefficients of the powers of $N_G$ involved with the large numbers are obtained as explicit functions of microscopic, measurable and recorded\cite{Tay:1991} experimental numerical values of these physical parameters.  In the next section of this paper, it will be shown that the theory developed in $A$ is consistent with the mathematical structure of the Friedman Cosmological models. A specific quantized oscillating model is obtained using the epoch dependent gravitation function $G(t^*\cos (\chi _G))$ and the mass of the universe, $M_U$ derived from quantum gravity theory in $A$.  

\section{The Friedman metric}
\setcounter{equation}{0}
\label{sec-ttln2}

The quantum theory for gravity developed in $A$ was formed from quantum theory based on concepts from Bohr's work, from Sommerfeld's  work\cite{Som:1916}, from a theory\cite{Gil:1996} for quantum coupling constants and a {\it time\/} dependent formula\cite{Gil:1999} for Newton's gravitation constant $G$. Thus here the term constant for $G$ is a misnomer. However, the two quantities radius and mass, of the universe come out of that theory together with the conclusion that the graviton has a rest mass negligibly greater than zero. An explanation for the rest mass value for the proton as being due to graviton kinematics also comes out of that theory. However, apart from the input of the formula for $G(t)$, no general relativity theory in the form of field equations or metric is used as input.
Thus the main question that arises is how does this all reflect on relativistic cosmology theory. It will be shown here that the quantum theory structure integrates perfectly with standard Friedman cosmology theory.

The standard cosmological Friedman metrics are of the form

\begin{eqnarray}
ds^2 & = &(cdt)^2 - R^2(t)\{\frac{dr^2}{1-kr^2}\nonumber\\
 & + & r^2 d\theta ^2 + r^2\sin ^2(\theta )d\phi ^2\}\label{1B01}
\end{eqnarray}
where $R(t)$ is the length dimensioned scale function of time, $t$ and $k=+1,0,-1$ is the curvature parameter with the three indicated allowable values. The rather simplistic meanings of these three cases are universes that are {\it closed\/}, {\it open and finite\/}, {\it open and infinite\/} respectively, See Rindler\cite{Rin:2001} page 367 for more detail about the parameter $k$. The scale factor, R(t), often called the expansion factor is the only function that needs to be found from the general relativity field equations to produce a cosmological theory. However, the form and physical structure of the theory generated depends on input to the field equation. The quantum theory of gravity from $A$ has given the main requirements for input into the field equations. They are the mass of the universe $M_U(N_G)$, at the present epoch as represented by  the key quantum number $N_G$ or by  $t^*$, the gravitation constant $G$ as a function $G(r'^*)$ of the radius $r'^*$ of a general proton's {\it gravitional orbit\/} and how this orbit with radius, $ r'^*$, depends on the radius of the universe, $r^*$. This input information is displayed next for easy reference
\begin{eqnarray}
G(t'^*) & = & \hbar ^2/(m^2_pm_e t'^*c)\label{1B02}\\
M_U(N_G)& = & \frac{N_G^2 m_e}{\cos (\chi_G(N_G))} \label{1B03}
\end{eqnarray}

The first move is to solve the general relativity field equations to find the two equations that are often called the {\it Friedman\/} equation and the {\it second\/} equation for $r(t)$ as a function of the time $t$. These equations have the forms (\ref{1B04}) and (\ref{1B05}),
\begin{eqnarray}
8\pi G\rho r^2/3 & = & \dot{r}^2 +(k -\Lambda r^2/3)c^2 \label{1B04}\\
 -8\pi GP'r/c^2 & = & 2\ddot{r} +\dot{r}^2/r +(k/r -\Lambda r)c^2, \nonumber\\
&  & \label{1B05}\\
P' & = & c^2\rho /3\label{1B041}
\end{eqnarray}
where the quantity $G$ on the left hand side of these equations is the usual constant gravitation constant, $G$. $k$ is the curvature parameter that can take on the values $-1,0,+1$ and $\Lambda$ is the cosmological constant. The last equation here (\ref{1B041}) is the assumption that the density function $\rho$ is capable of exerting the pressure $P'$. The three equations (\ref{1B04}), (\ref{1B05}) and (\ref{1B041}) form the usual basis for the non-quantal cosmologies that are discussed in great detail, in Rindler's book\cite{Rin:2001}. Other assumptions can be made at this point which will render these equation more suitable for describing a cosmology based on the quantum gravity theory which was outlined in $A$. This will be dealt with in the next section.
\section{Expansion Process}
\setcounter{equation}{0}
\label{sec-expp} In the quantum system developed earlier, mass is generated over time {\it inside\/} the expanding universe.
In $A$, I showed how {\it conservation of energy\/} for the quantum expansion process can be seen to hold by regarding any increase $\delta M$ of mass within the outer expanding boundary over some time $\delta t$, say, as arising from mass enveloped from outside the expanding boundary over the time $\delta t$, $\delta M = c\delta t 4\pi R^2\rho $ where $\rho$ is the mass density just outside the expanding boundary. It follows that the process occurring in this system is not to be thought of as {\it continuous creation\/}. It is necessary for this development that there are two mass densities involved $\rho _i$ and $\rho _o$, $\rho _i$  inside the moving spherical boundary and $\rho _o$ outside the moving boundary. In principle they can exert positive pressures $P_i$, $P_o$ away from their regions of location, if they are physically suitably constituted by formulae such as $P_i=c^2\rho _i/3$ and $P_o=c^2\rho _o/3$. Thus the effective pressure exerted outward from the expanding universe on to its expanding boundary will be $ P_{o,e}=P_i -P_o$ if the contributions from both densities are taken into account. The mass density distribution, $\rho _o$ outside the expanding universe, I shall call the universe's mass halo. It is not observationally accessible from within the universe.  The expanding universe consumes it own halo. On the basis of these remarks about inside and outside mass densities, we can give a clear qualitative explanation of the nature of the expansion process. It can be regarded as a spherically expanding change of state or phase transition in which the material outside the moving spherical boundary in state $S_o$ is continuously consumed to reappear within the boundary in state $S_i$. The state inside and the state outside having characteristics reflected in the form of equation of state. I shall take it to be the case that the external state is that of a continuous mass density satisfying $P_o=c^2\rho _o/3$ and the internal state is that of a {\it dust\/} distribution with a mass density $\rho _i = \rho _o$, the same as the external density at the boundary but unable to exert pressure, $ P_i = 0$. The change of state does {\it not\/} involve a change of density. Thus the total effective outwards pressure will be the negative pressure $P_{o,e} = - P_o$ so that we need to replace equations ((\ref{1B04})), (\ref{1B05}) and (\ref{1B041}) with the prime on $P'$ now dropped,    
\begin{eqnarray}
8\pi G\rho r^2/3 & = & \dot{r}^2 +(k -\Lambda r^2/3)c^2 \label{1B042}\\
 -8\pi GPr/c^2 & = & 2\ddot{r} +\dot{r}^2/r +(k/r -\Lambda r)c^2, \nonumber\\
&  & \label{1B051}\\
P & = & - c^2\rho /3\label{1B043}
\end{eqnarray}
\section{The Quantum Model}
\setcounter{equation}{0}
\label{sec-quam}
I shall confine the discussion in this paper to the cases when the cosmological constant is negative, that is $\Lambda = -|\Lambda |$. Using a radius variable version of the gravitation constant, $G(r')\leftarrow G(r'/c)$, the mass of the universe $M_U = \rho 4 \pi r^3/3$ and the quantization projection formula, (\ref{1B07}) for $R$, the quantity $8 \pi G(R\cos (\chi))\rho R^2/3$ on the left hand side of (\ref{1B04}) can be expressed in the form,
\begin{eqnarray}
8 \pi G(R\cos (\chi))\rho R^2/3 & = & 2(c\cos^2 (\chi))^2.\label{1B06}\\
R\cos (\chi)& = & N_G l''_p.\label{1B07}
\end{eqnarray}

Using (\ref{1B06}) in the Friedman equation (\ref{1B04}), that equation becomes
\begin{eqnarray}
2(c\cos^2 (\chi))^2 & = & \dot{r}^2 +(k -\Lambda r^2/3)c^2 \label{1B08}\\
 -8\pi GPr/c^2 & = & 2\ddot{r} +\dot{r}^2/r +(k/r -\Lambda r)c^2, \nonumber\\
 &  & \label{1B09}
\end{eqnarray}

or
\begin{eqnarray}
\dot{r}^2 & = &   (a(k, \chi _G(N_G))  + br^2)c^2\label{1B10}\\
a(k, \chi _G(N_G)) & = & (2\cos ^4 (\chi _G(N_G) ) -k) \label{1B11}\\
b & = & \Lambda /3 = -|\Lambda |/3\label{1B12}
\end{eqnarray}

Equation (\ref{1B10}) can be written in a form suitable for integration as

\begin{eqnarray}
 \frac{cdt}{dr} & = & \pm 1/( a(k, \chi _G(N_G)) + br^2)^{1/2}\label{1B13}\\
\int_{t_0}^{t_R}cdt& = & \pm\int _{R_0}^{R} \frac{dr}{( a(k, \chi _G(N_G)) + br^2)^{1/2}}\nonumber\\
\label{1B14}
\end{eqnarray}

The function of $r$, $ a(k, \chi _G(N_G)) $, in the denominator of equation (\ref{1B13}) makes for some difficulty in evaluating the integral because it is not a simple function. However, for larger values of $r$, $ a(k, \chi _G(N_G)) $ is indistinguishable from the pure constant $a(k,0)$ because the angle $\chi _G(N_G)$ is very small for large $r$  so that the integration is easily performed in that case to give a relation between $r$ and $t$ which is physically very good provided we do not interpret it for the smaller values of $r$. This is the course that will be taken in this article while the more involved situation for smaller values of $r$ will be dealt with in the next section. Thus under this restriction
our integral becomes 
\begin{eqnarray}
\int_{t_0}^{t_R}cdt& = & \pm\int _{R_0}^{R} \frac{dr}{(a(k,0) + br^2)^{1/2}}\label{1B15}\\
a(k,0)  & = & 2-k\label{1B16}\\
b & = & -|\Lambda /3|\label{1B17}\\
k& = &\ (-1,0,+1)\label{1B18}
\end{eqnarray}
which is a standard form with the possible values from integration,
\begin{eqnarray}
R(t)& = & a(k,0)^{1/2} R_\Lambda\sin (\pm c(t-t_0)/R_\Lambda +\nonumber\\
 & & \sin ^{-1}(R_0/( a(k,0)^{1/2} R_\Lambda)))\label{1B19}\\
R_\Lambda & = & |3/\Lambda|^{1/2}\label{1B20}\\
R(t) & = & R_0,\ when\ t=t_0\label{1B21}
\end{eqnarray}
As this solution is rigorously cyclical, while not forgetting that the true physical case would involve the function $a(k,\chi _G(N_G))$ rather than $a(k,0)$, we might just as well discuss this for the simplest case $R(0) = R_0$,
$a(1,0)=1$ and with only the plus sign in the sine function. Then
\begin{eqnarray}
R(t)& = & R_\Lambda\sin (ct/R_\Lambda) \label{1B22}\\
\dot R(t)& = & c\cos (ct/R_\Lambda)\label{1B220}\\
R_\Lambda & = & |3/\Lambda|^{1/2}\label{1B23}\\
R(t) & = & 0,\ when\ t= t_0 = 0\label{1B24}
\end{eqnarray}
and we see that everything depends on the value assigned to $\Lambda$ as this will determine the maximum value for $R(t)$ which is given by $R_\Lambda$ through (\ref{1B23}) according to (\ref{1B22}). Thus the cosmological constant assumes prime importance for this quantum cosmology but it still remains a numerical value that does not come out of the theory but remains rather a value that needs to be found from experiment or observation and then input into the theoretical construction. The value of $\Lambda$ in this construction is directly related to the ultimate radius of the universe $R_\Lambda$. A list of values for important parameters can be found in the Wheeler book\cite{{Mis:1973}} on page 738, (Box 27.4). I shall use some of this information to determine the viability of this quantum model as representing the physically observed or assumed values for describing the universe after considering the second Friedman equation and some physically measurable characteristics.

 If we take the difference of (\ref{1B04}) and (\ref{1B05}) and take (\ref{1B042}) into account we obtain successively,

\begin{eqnarray}
\ddot R(t)/R & = & -4\pi G(\rho /3 + P/c^2) -|\Lambda|c^2/3\nonumber\\
&  & \label{1B25}\\
& = & -|\Lambda|c^2/3\label{1B26}\\
& = &  - c^2/R_\Lambda ^2\label{1B27}\\
\ddot R(t) & = &  - \omega ^2R(t)\label{1B28}\\
\omega & = &  c/R_\Lambda = c |3/\Lambda|^{-1/2}.\label{1B29}
\end{eqnarray}
Hence we have a simple harmonic universe with angular frequency parameter $\omega$ determined by the cosmological constant $\Lambda$ through (\ref{1B28}).
Thus the acceleration of this universe is negative for positive $R(t)$ and positive for negative $R(t)$. The values of the three functions of cosmic time $H,\Omega,q$, Hubble's constant, the dimensionless density parameter and the dimensionless deceleration parameter that are used to test theory against observation are,
\begin{eqnarray}
H & = & \dot R/R = c \cot (ct/R_\Lambda)/R_\Lambda\label{1B30}\\
\Omega & = & 8\pi G\rho/(3H^2)= 2 (\cos (\chi _G)R_\Lambda /R)^2\nonumber\\
&  & \label{1B31}\\
q & = & -\ddot R/(RH^2)=\tan ^2(ct/R_\Lambda).\label{1B32}
\end{eqnarray}

The quantum model cosmology could not be more simple than that described by the function $R(t)$ above where the cosmological constant $\Lambda$ is a free input parameter not determined through the present theory. Thus all that remains to be shown is that it can be assigned a value that gives good agreement for the values of other cosmological quantities that have been measured at the present epoch. The  most important of these is the cosmological present day radius of the universe which is derived in the quantum gravitation theory to be given by  $ct^* = 2.18c\times 10^{17} m$. However the present day age of the universe can not now be taken to be $t^* = 2.18\times 10^{17} s $ because the relation between time and radius has now been shown to have the more involved form (\ref{1B22}). In order to make this distinction between possible representations of ages of the universe let us now denote the cosmological quantum age or epoch by $t^\dagger$ and regard $t^*$ as a {\it formal age\/} or just the time a light ray at velocity c would take to transverse the radius if that kept constant at the value, $r^*$. The most important physical quantities of the present day  universe at their present day epoch values are, Hubble's constant $H(t^\dagger)$ and $R(t^\dagger)$ given below with their measured values. We can solve the two equations (\ref{1B33}) and (\ref{1B34}) to obtain the values for $R_\Lambda$, $\Lambda$ and $t^\dagger$ given at equations (\ref{1B35}), (\ref{1B36}) and (\ref{1B37}).  
\begin{eqnarray}
R(t^\dagger) & = & r^* = 2.18c\times 10^{17}= 6.535\times 10^{25}\  m\nonumber\\
& &\label{1B33}\\
H(t^\dagger) & = & \dot R/R = c \cot (ct^\dagger/R_\Lambda)/R_\Lambda\nonumber\\
& = & 2.1056\times 10^{-18}\ s^{-1}\label{1B34}\\
t^\dagger & = & 2.684\times 10^{17}\ s\label{1B35}\\
R_\Lambda & = & 7.356\times 10^{25}\ m\label{1B36}\\
\Lambda & = & 5.544 \times 10^{-52}\  m^{-2}\label{1B37}\\
t_\Lambda & = & R_\Lambda\pi/(2c)= 3.864\times 10^{17}\ s\label{1B38}\\
R_\Lambda /R(t^\dagger) & = & 1.126\label{1.39}\\
t_\Lambda /t^\dagger & = & 1.436\label{1B40}\\
T_U & = &  2\pi/\omega = 4.9\times 10^{10}\ yr\label{1B41}
\end{eqnarray}

The simple formula (\ref{1B22}) together with the value for the constant maximum radius $R_\Lambda$ gives a model for a {\it quantum\/} cosmology that agrees with the experimental information available with high accuracy. The numerical value for $R_\Lambda$ depends only on the value assumed for the cosmological constant $\Lambda$. Thus here there is no question of whether the cosmological constant is important or not, its value and existence is fundamental and crucial to this quantum version of cosmology. The version for the model developed here has not taken into account the mostly near zero angle $\chi_G$ which will have more importance for small values of epoch. This will be dealt with in the next section. However, the approximate model that has been developed here does give all the essential feature of a quantum model for larger epoch values. The main feature is that the model is cyclical or simple harmonic with a period of approximately $ 4.9\times 10^{10}$ years. It is interesting that the that the ratio of maximum radius to radius now is $R_\Lambda /R(t^\dagger )=1.126 $ whereas the ratio time at maximum radius to time now is $t_\Lambda /t^\dagger = 1.436$. This means that radius-wise now we are relatively quit near to the turn round point when contraction begins but time-wise that turn round is some way in the future.

It has been demonstrated that  that the quantum theory for gravity obtained in $A$ is consistent with the mathematical structure of the Friedman Cosmological models. The specific quantized oscillating model was obtained using the epoch dependent gravitation function $G(t^*\cos (\chi _G))$ and the mass of the universe, $M_U$ derived from quantum theory. However, that model was restricted to an unspecified range of radius values vaguely called large. This was necessary because of problems with the evaluation of a key integral. In the next section this restriction to range is lifted and the angle $\chi_G$ is taken to be nonzero to give a model valid over a very large range of values for the epoch changing universe radius.  

\section{A Second Quantum\\Friedman Cosmology}
\setcounter{equation}{0}
\label{sec-ttln3}

The quantum theory for gravity developed in $A$ was formed from quantum theory based on concepts from Bohr's work, from Sommerfeld's  work, from a theory for quantum coupling constants and a {\it time\/} dependent formula for Newton's gravitation constant $G$. The author's theory for the quantum coupling constants depends on quantum and relativity concepts introduced via an idea called {\it projection quantization\/}. Projection quantization is more familiar in quantum studies in relation to angular momentum. However, it is used in this work in relation to the ordinary geometry of spatial lengths and spatial angles and effectively supplies a bridge between length quantization and relativity length contraction. It comes into the quantization of gravitation via the formula,

\begin{eqnarray}
r^* \cos (\chi_G(N_G))& = & N_G l''_p\label{9.00}\\
 l''_p & = & l_p/\cos^2 (\chi_G(n_G))\label{9.01}
\end{eqnarray}
$r^*$ is to be identified with the present day radius of the universe and $l_p$ is the crossed Compton wave length of the proton $\hbar /(m_pc)$. The quantity $ r'^* = r^*\cos (\chi_G(N_G))$ is the radius of the very large radius gravitational orbit that any proton exists on as a consequence of being part of the universe and subject to the gravitational influence of the rest of the universe. The gravitational orbit of the general proton under the gravitational attraction of the rest of the universe is exactly analogous to the orbit of an electron in the electromagnetic orbit of a hydrogen-like atom under the Coulomb potential of its nucleus. The analogue of the quantity $\cos (\chi_G(N_G)) = N_G\alpha _G$ is the quantum electromagnetic quantity $\cos (\chi _{137})= 137\alpha$ in the case of hydrogen$_{137}$. This is all explained in great detail in $A$. In the previous section, the Friedman cosmological equations from general relativity were used to derive a {\it quantum\/} cosmological model and it was shown that the numerical value of the cosmological constant could be chosen together with identifications involving the measured value of $G$ to give very accurate model agreement with present day astronomical measurement. The steps involved in deriving this quantum model universe depended on performing an integration of the Friedman equations which was carried out under the assumption that $\cos (\chi_G(N_G))=1$. This assumption is in fact very accurately correct and increasingly so for values of $N_G > 137$ as can be seen by using the general coupling constant formula which gives $\cos (\chi) = 137 \alpha \approx 1 - 0.000263$.
Thus the first and second models could be said to be reliable for $N_G >137$ or equivalently for $r^* >137 l_p \approx 2.9 \times 10^{-14} < 10\times 10^{-14} = 10^{-13}$ meters. It follows that for a range of radius values ranging from the subatomic $10^{-13}\  m$ to the ultimate radius of the universe the first model is valid. However, it still remains very interesting to explore just what the effect on the model would be by taking the projection cosine  $\cos (\chi_G(n_G))$ not equal to unity but rather taking it to have an acceptable dependence on $r^*$ and then performing the integration in that case. It is also desirable to carry through the more general case just in case there are unexpected nasty surprises such as singularities that would have been missed in the simple case. It turns out that this can be done as will now be shown. Firstly, for ease of reference here at (\ref{9.03}) is the integral that has to be performed,
\begin{eqnarray}
 \frac{cdt}{dr} & = & \pm \frac{1}{( a(k, \chi _G(N_G)) + br^2)^{1/2}}\label{9.02}\\
\int_{t_0}^{t_R}cdt& = & \pm\int _{R_0}^{R} \frac{dr}{( a(k, \chi _G(N_G)) + br^2)^{1/2}}\nonumber\\
\label{9.03}\\ 
a(k, \chi _G(N_G)) & = & (2\cos ^4 ( \chi _G(N_G) ) -k),\nonumber\\
a(k, \chi _G(N_G)) & \approx & (2\cos ^2 (\chi _G(N_G) ) -k)\label{9.04}
\end{eqnarray}
as $\chi _G(N_G) )$ is very small. Using the power 2 rather than 4 only makes a very small difference to the constant function $B(R_\Lambda)$ that appears at formula ($\ref{9.14}$). 
I shall restrict the discussion to what is called the closed universe case $k=1$ so that the first need is to find an $r$ variable function to replace the quantity $a(1, \chi _G(N_G))$. This can be achieved as follows,
\begin{eqnarray}
a(1, \chi _G(N_G))& \approx & (2\cos ^2 ( \chi _G(N_G) ) -1)\nonumber\\
&\approx& \cos (2\chi _G(N_G)) \label{9.05}\\
& \approx & \cos (2\pi/N_G)\approx 1 - 2(\pi/N_G)^2\nonumber\\
& & \label{9.06}\\
N_G &\approx & r^*/l_p\label{9.07}\\
a(1, \chi _G(N_G))& \approx & 1 - 2 (\pi l_p/r^*)^2. \label{9.08}
\end{eqnarray}
Four approximation steps are used in getting to the final form (\ref{9.08}). Each of these approximations are accurate for all $N_G$ and very accurate indeed for $N_G>137$. Thus one can have great confidence in the final formula (\ref{9.08}) as usable in the integral \ref{9.03}. The integral that now has to be performed is
\begin{eqnarray}
 \int_{t_0}^{t_R}cdt& = & \pm\int _{R_0}^{R} \frac{dr}{( 1 - 2 (\pi l_p/r)^2 + br^2)^{1/2}}.\nonumber\\
\label{9.09}
\end{eqnarray}

The integral (\ref{9.09}) can be carried through after multiplying numerator and denominator by $2r$ and using the transformation (\ref{9.10}) to give the result (\ref{9.11}) together with definitions (\ref{9.12}) $\rightarrow$ (\ref{9.16}).
\begin{eqnarray}
r(t) & \rightarrow & w(t) = r^2(t) + 1/(2b)\label{9.10}\\
r^2(t) & = & R_\Lambda B(R_\Lambda)\sin\phi(t) + R_\Lambda ^2/2\nonumber\\
\label{9.11}\\
\phi(t) & = & \frac{\pm 2ct}{R_\Lambda} + \phi_0(R_\Lambda)\label{9.12}\\
\phi _0(R_\Lambda) & = & \sin ^{-1}(-R_\Lambda/(2B(R_\Lambda)))\label{9.13}\\
B(R_\Lambda) & = & ((R_\Lambda/2)^2 - 2(\pi l_p)^2)^{1/2} \label{9.14}\\
R_\Lambda & = & (|\Lambda|/3)^{-1/2}\label{9.15}\\
\Lambda & = & 3b\ <\ 0\label{9.16}
\end{eqnarray}
The value for Hubble's constant at epoch $t$ is given by
\begin{eqnarray}
 H(t)& = & \frac{\dot r(t)}{r(t)}= \frac{c B(R_\Lambda)\ cos \phi(t)}{r^2(t)}\label{9.17}
\end{eqnarray} 
Let us now eliminate the sine and cosine functions between (\ref{9.11}) and (\ref{9.17}) using $\cos ^2(\phi) + \sin ^2 (\phi) -1 =0$ to obtain the zero valued function FQ(t) of $t$,
\begin{eqnarray}
 FQ(t) & = & (r^2(t) -R_\Lambda ^2/2)^2c^2 + (H(t)r^2(t)R_\Lambda)^2\nonumber\\
& & - (R_\Lambda B(R_\Lambda))^2\label{9.18}
\end{eqnarray}
The numerical value of the radius of the universe at epoch now, $t^\dagger$, identified from the quantum theory for the gravitation constant, $r^*(t^\dagger) =r(t^\dagger)$, and the experimental value for Hubble's constant $H(t^\dagger)$ at time $t^\dagger$ can be used in this equation to give an equation, $FQ(t^\dagger)=0$ from which the value of $R_\Lambda$ and hence the value of $\Lambda$ can be obtained.
 \begin{eqnarray}
 FQ(t^\dagger) & = & (r^2(t^\dagger) -R_\Lambda ^2/2)^2c^2 + (H(t^\dagger)r^2(t^\dagger)R_\Lambda)^2\nonumber\\
& & - (R_\Lambda B(R_\Lambda))^2\label{9.19}\\
r(t^\dagger)  & = & 6.535\times 10^{25}\ m\label{9.20}\\
H(t^\dagger) & = & 2.1056\times 10^{-18}\ s^{-1}\label{9.21}\\
R_\Lambda &  = & 7.355566 \times 10^{25}\ m\label{9.22}\\
\Lambda &  = & -5.54484054 \times 10^{-52}\ m^{-2}\label{9.23}
\end{eqnarray}
To find the numerical values of the time now, $t^\dagger;\  r^* = r(t^\dagger)$, and the time at maximum radius $t_{max}$ we can use the formulae (\ref{9.24}) and (\ref{9.25}),

\begin{eqnarray}
 t^\dagger & = & R_\Lambda(\sin ^{-1}(\frac{r^{*2} -R_\Lambda ^2/2}{R_\Lambda B(R_\Lambda)})-\phi_0(R_\Lambda))/(2c)\nonumber\\
& & \label{9.24}\\
t_{max}  & = & R_\Lambda(\pi/2 -\phi _0(R_\Lambda))/(2c)\label{9.25}\\
r_{max}  & = & r(t_{max})=7.355566\times 10^{25}\label{9.26}\\
H(t^\dagger) & = & 2.1056\times 10^{-18}\ s^{-1}\label{9.27}\\
R_\Lambda &  = & 7.355566 \times 10^{25}\ m\label{9.28}\\
\Lambda &  = & -5.54484054 \times 10^{-52}\ m^{-2}\label{9.29}
\end{eqnarray}

Finally here is a list of the main numerical values and ratios arising from this second quantum cosmology model each one proceeded with the corresponding value obtained in the original $\chi _G(N_G)=0$ model and distinguished by a prime.

\begin{eqnarray}
\Lambda ' & = & -5.5438\dots\times 10^{-52}\ m^{-2} \label{9.30}\\
\Lambda   & = & -5.5448\dots\times 10^{-52}\ m^{-2}\label{9.31}\\
 R_\Lambda ' & = & 7.356244\dots\times 10^{25}\ m\label{9.32}\\
 R_\Lambda   & = & 7.355566\dots \times 10^{25}\ m\label{9.33}\\
 r_{max}' & = & R_\Lambda ' \ m\label{9.34}\\
 r_{max}  & = & R_\Lambda \ m\label{9.35}\\
 t_{max}' & = & 3.85438724\dots\times 10^{17}\ s \label{9.36}\\
 t_{max}  & = & 3.85403596\dots\times 10^{17}\ s\label{9.37}\\
 r'_{max}/r'_{\dagger} & = & 1.1255867\dots\label{9.38}\\
 r_{max}/r_{\dagger}   & = & 1.1255648\dots\label{9.39}\\
 t'_{max}/t'_{\dagger} & = & 1.4359549\dots\label{9.40}\\
 t_{max}/t_{\dagger}   & = & 1.4359054\dots\label{9.41}\\
 T'_U & = & 4.8888727\dots\times 10^{10}\ yrs\label{9.42}\\
 T_U  & = & 2.4421080\dots\times 10^{10}\ yrs\label{9.43}\\
 t'^\dagger & = & 2.68419786\dots\times 10^{17}\ s\label{9.44}\\
 t^\dagger  & = & 2.68404281\dots\times 10^{17}\ s\label{9.45}
\end{eqnarray}

\section{Conclusions}
The formula (\ref{9.11}) together with the value for the constant maximum radius $R_\Lambda$ gives a model for a {\it quantum\/} cosmology that agrees with the experimental data to high accuracy. $R_\Lambda$ is calculated from the theory and in turn determines the value of the cosmological constant, $\Lambda$. The second model {\it has\/} taken into account the mostly near zero projection angle, $\chi_G$, in contrast to the first model. This more in depth study has not shown up any dangerous physical or mathematical complications such as {\it singularities}. Importantly, there is no big bang. The main feature is that the model is smoothly cyclical or simple harmonic with a period of approximately $2.442108\times 10^{10}$ years in contrast with the $4.48888727\times 10^{10}$ years from the first model, apparently twice as long to complete a full cycle. This is in fact not an indication of a great difference between the two models. It partly arises because we are not comparing like with like. The radius in the first model depends on a sine function whereas in the second model the radius squared depend on a similar sine function at twice the frequency. Effectively, this means that the second sine wave appears as the modulus of the first sine wave so that over the period the negative values of the first sine wave are sign inverted. Thus one complete cycle of the first model corresponds to two complete cycles of the second model. One may view the second model as having the physical advantage of only giving positive values for $r$ over its complete cycle unlike the situation for the first model that allows negative $r$ for its second half cycle. This factor $2$ is a consequence of different mathematical representations of two similar systems. The physical difference associated with this aspect is more reflected in the fact that this factor is not exactly $2$. The observation that the ratio of maximum radius to radius now is $r_{max}/r(t^\dagger )=1.1255648 $ whereas the ratio time at maximum radius to time now is $t_{max}/t^\dagger = 1.4359054$ is much the same as in the first model with the same implication that radius-wise now we are relatively quit near to the turn round point when contraction begins but time-wise that turn round is some way in the future. The rest of the numerical comparisons in list (\ref{9.30})$\rightarrow $ (\ref{9.45}) show close {\it similarity\/} for the two models. However, one should recognise that the small differences in the decimal parts of the various parameter tend rather to hide the great differences in value that are really involved. For example, the numerical difference between $t'^\dagger$ and $t^\dagger$ is $1.5505\times 10^{13}\ s\approx 5463$  {\it life-spans\/}.  From a human perspective this is a long time. Perhaps the most interesting result from these quantum models is that there is no {\it Big Bang\/} and conservation of energy still rules.  

\section{Acknowledgements}
I am greatly indebted to Professors Clive Kilmister and Wolfgang Rindler for help, encouragement and inspiration over many years.

\end{document}